\documentclass[floatfix,amsmath,amssymb,amsfonts,bbm,pra]{revtex4}
\usepackage{color}
\usepackage{amsmath}
\usepackage{amsfonts}
\usepackage{bbm}
\usepackage{verbatim}
\usepackage{hyperref}
\usepackage{graphicx}

\input{Qcircuit}

\newcommand{\bracket}[2]{\langle #1|#2\rangle}

\newcommand{\UU}{\mathcal{U}}

\newcommand{\be}{\begin{eqnarray}}
\newcommand{\ee}{\end{eqnarray}}

\newcommand{\Tr}{\text{Tr}}

\begin{document}

\author{L. Sheridan}
\affiliation{Institute for Quantum Computing, University of Waterloo, Waterloo, ON, N2L 3G1, Canada}
\author{D. Maslov}
\affiliation{Institute for Quantum Computing, University of Waterloo, Waterloo, ON, N2L 3G1, Canada}
\affiliation{National Science Foundation, 4201 Wilson Blvd, Arlington, VA, 22230, USA}
\author{M. Mosca}
\affiliation{Institute for Quantum Computing, University of Waterloo, Waterloo, ON, N2L 3G1, Canada}
\affiliation{Perimeter Institute for Theoretical Physics, 31 Caroline Street North, Waterloo, ON, N2L 2Y5, Canada}
\affiliation{St. Jerome's University, Waterloo, ON N2L 3G3, Canada}
\affiliation{Department of Combinatorics \& Optimization, University of Waterloo, ON N2L 3G1}

\title{Approximating Fractional Time Quantum Evolution}

\date{October 20, 2008}

\begin{abstract}
An algorithm is presented for approximating arbitrary powers of a black box unitary operation, $\mathcal{U}^t$, where $t$ is a real number, and $\mathcal{U}$ is a black box implementing an unknown unitary.  The complexity of this algorithm is calculated in terms of the number of calls to the black box, the errors in the approximation, and a certain `gap' parameter.  For general $\mathcal{U}$ and large $t$, one should apply $\mathcal{U}$ a total of $\lfloor t \rfloor$ times followed by our procedure for approximating the fractional power $\mathcal{U}^{t-\lfloor t \rfloor}$. An example is also given where for large integers $t$ this method is more efficient than direct application of $t$ copies of $\mathcal{U}$.  Further applications and related algorithms are also discussed.
\end{abstract}

\maketitle

\section{Introduction}
\label{sec:intro}

Suppose an $n$-qubit unitary $\UU$ is implemented by evolving (or simulating the evolution of) a {\it time-independent} Hamiltonian $H$ for a period of 
time $\tau=1$, that is, $U = e^{-i H}$. Then for any $t\in\mathbb{R}_{+}$, one can implement $\UU^t$ by simply evolving the Hamiltonian for a period of 
time $t$.  For example, if $t= \frac{1}{2}$, then a square root of $\UU$, $e^{-i\frac{1}{2} H}$,  could be implemented in this way, and in such model of computation the cost would
be half of the cost of implementing $\UU$.

In our work, we question what can be done if $\UU$ is realized in some other way, such as a non-trivial sequence of time-dependent Hamiltonians, or a 
quantum circuit.  In other words, consider the situation when $\UU$ is given in the form of a black box. The goal is to implement real valued powers $t$ 
of this unitary operation by making use of the multiple copies of the black box implementing $\UU$.  The complexity of such a procedure is measured in terms 
of the total number of calls to the unitary.

It is possible to find the $t^{\textrm{th}}$ power of an unknown unitary by first performing a sufficiently precise
  complete quantum process tomography of a $2^n \times 2^n$ dimensional unitary $\UU$, which uses
 $O(4^{n})$ calls to the unitary with various input states~\cite{NC,NCb} and measurements to achieve $\UU$ with constant precision.  The exponential scaling with $n$ is necessary.
In particular, a lower bound on the number of calls to the unitary can be derived from constructing an $\epsilon-$net over unitaries~\footnote{We thank P. Hayden for this observation.}.  If the space of unitary operations is divided up into balls of radius $\epsilon$ then the total number of unitaries that can be specified up to this resolution is $\left(\frac{c}{\epsilon}\right)^{4^{n}}$, where $c$ is a constant.  To specify each of these requires $4^{n} \log(\frac{c}{\epsilon})$ bits.  To discover information about $\UU$ one might supply states to the unitary tensored with the identity operation of the same dimension (a larger dimension does not help) and perform measurements on the output states.  The states will have dimension $4^{n}$, and by Holevo's bound one could hope to obtain no more than $2n$ bits of information for each of these calls to $\UU$ (in general we will get less than this).  Therefore, it is not possible by tomography to use fewer than $\Omega\left(\frac{4^n}{2n} \log (\frac{c}{\epsilon})\right)$ calls to $\UU$.  Even if we allow for error in some fraction  $\delta$ of the basis states, this still cannot reduce the required number of calls to any function polynomial in $n$~\cite{aar:learn}.

In this paper, we report on a more efficient approach to implementing
powers.  In particular, we present an algorithm for approximating with error
$\epsilon$ (using the trace norm) any constant power of an unknown unitary
using only $O(\frac{1}{\epsilon} \log \frac{1}{\epsilon})$ calls to the
unitary itself, assuming a certain `gap' parameter $g$ we explain later
satisfies $g \in \Omega(\epsilon)$ (otherwise, the complexity is $O(\frac{1}{g} \log
\frac{1}{\epsilon})$ calls; see Section~\ref{optimality} for more detail).  Note that this complexity is independent of the number of qubits $n$.

There has been some relevant prior work concentrating on the relation between discrete and continuous oracles for various problems.  Farhi and 
Gutmann~\cite{FG} introduced the concept of (continuous) Hamiltonian black box oracles for quantum computing. Ioannou~\cite{ioannou}, and Roland and 
Cerf~\cite{RC} consider the problem of simulating a Hamiltonian for Grover search~\cite{Gsearch} on a discrete computer.  Mochon~\cite{Mochon} extends this to a more 
general setting where he is concerned with finding lower bounds for discrete oracle problems by considering them in the Hamiltonian setting and mapping them to the problem of finding geodesics in manifolds.  In particular, he considers one-item Grover search and oracle interrogation, highlighting the case of 
computing the XOR function on a hidden bit string.  In these cases, he is able to exploit symmetries in the oracle problems to solve what is otherwise a 
very difficult problem.
  Our focus is different.  Firstly, we do not allow for an oracle to be applied a fractional amount of time (instead, we are 
concerned with simulation of fractional oracles in the case when an oracle is only allowed to be applied an integer number of times).  
Secondly, we present an algorithm, with our focus on (constructive) upper bounds.  Thirdly, we do not make symmetry 
assumptions on our oracles. Finally, we do not restrict the eigenvalues of these unitaries to $\pm 1$ as in the Grover search and
XOR case, or the work in \cite{CGMSY08} which shows how to simulate a general continuous {\it Boolean} oracle evolving for a total of time T with $O\left(\frac{T \log T}{\log\log T}\right)$ discrete oracle queries.

The remainder of the paper is organized as follows.  In Section~\ref{sec:algo} we describe the basic construction for efficient computation of the powers 
of $\UU$, discuss our assumptions and constraints, give the complexity of the computation, and calculate the precision of our approximate solution.  The 
full calculation of these results is included in~\ref{sec:complerrors}.  An example of a special case in which $t$ is a large positive integer and our 
algorithm is more efficient than direct application of the $t$ copies of the unitary is given in Section~\ref{sec:example}.  Related applications of our 
algorithm to computing fractional Fourier transform and noise filtering are explored in Section~\ref{sec:ft}.  Finally, we discuss the implications of this work in the concluding section.

\section{The Algorithm} \label{sec:algo}

In this section, we describe our basic construction.  For any unitary $\UU$ on a finite dimensional state space, consider its spectral decomposition $\UU 
= P \Lambda P^\dag$, where $P$ is a unitary matrix composed of the eigenvectors of $\UU$, and $\Lambda$ is a diagonal matrix containing the eigenvalues of 
$\UU$.  For such a decomposition, powers of $\UU$ may be computed as follows:
\begin{equation}
\UU^t = P \Lambda^t P^\dag.
\end{equation}
A key observation is that in the quantum case one does not need to actually implement the basis change operators $P$ and $P^\dag$ in order to exploit the 
above feature of the spectral decomposition.

For now, let us assume that $t$ is real number between 0 and 1.

\begin{figure}[h!]
\begin{center}
\includegraphics[width=6in]{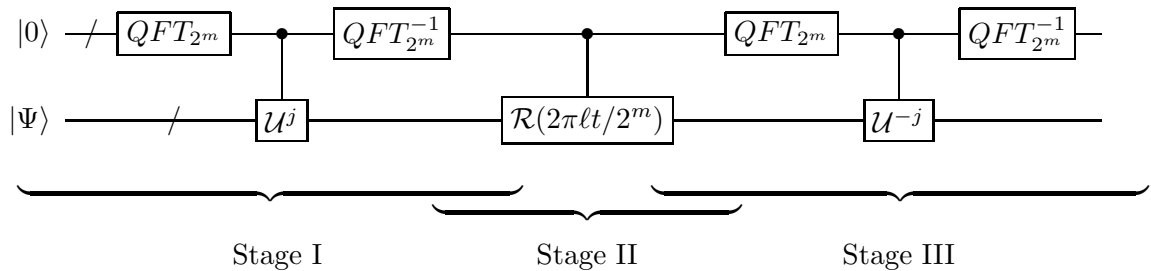}
\caption{The quantum circuit diagram outlining the implementation of the algorithm that raises an unknown unitary $\UU$ to some power $t$. Stage I performs eigenvalue estimation, where value $l$ in the first register corresponds to the eigenvalue parameter estimate $\lambda = l/2^m$. Stage II then uses this eigenvalue estimate to induce a corresponding phase shift $e^{2 \pi i lt/2^m}$.  Stage III uncomputes the eigenvalue estimation step. In the case that the eigenvalue parameters $\lambda_k$ are of the form $l_k/2^m$, the eigenvalue estimation is exact and the uncomputation step returns the control register back to the initial all-zeroes state.  In general, when we do not have this assumption on the form of the eigenvalues, we perform a slightly more complicated eigenvalue estimation algorithm that gives an estimate with error at most $1/2^m$ with probability in $1-O(1/2^{m})$.}
\label{AlgCircDiag}
\end{center}
\end{figure}

Our algorithm is described in three stages.  In the first stage, approximations to the eigenvalues of $\UU$ are calculated using an eigenvalue estimation
algorithm.  Let us initially assume that the eigenvalues are of the form $e^{2 \pi i \lambda_k}$ with $\lambda_k = \frac{\ell_k}{2^m}$ for some integer 
$\ell_k \in \{0,1,\ldots, 2^m -1\}$ (so that we can initially ignore the effect of precision errors).  In the second stage, phase shifts are applied to 
the eigenstates of $\UU$.  The third stage uncomputes the eigenvalue estimation.  Refer to Figure~\ref{AlgCircDiag} for details.

Suppose $\ket{\Psi}$ is a quantum state on which we wish to perform the transformation $\UU^t$.  Note that any state $\ket{\Psi}$ is a superposition of 
eigenvectors, $\ket{\psi_k}$, of $\UU$, $\ket{\Psi} = \sum_{k} \alpha_k \ket{\psi_k}$.  Prepare $m$ ancilla qubits in the ``all-zeros'' state, where $m$ 
is a user chosen parameter that characterizes the complexity of and errors in our algorithm; its role is discussed in more detail in \ref{sec:complerrors}.  
These ancilla are input to the Quantum Fourier Transform (QFT) circuit~\cite{KLM} (or equivalently, since applied to the all-zeros state, a circuit 
consisting of only Hadamard operations on each qubit) and later they are used as the controls for controlled-$\UU$ operators, while the state $\ket{\Psi}$ 
is the target.  The notation $c-\UU^j$ denotes the operation $\ket{j}\ket{\phi} \mapsto \ket{j}\UU^j \ket{\phi}$.  Using the eigenstate expansion of 
$\ket{\Psi}$ in the target, it can be shown that such a controlled operation causes a phase kick-back to the control register, so that the total state 
becomes $\sum_k \sum_j \alpha_k e^{2\pi i \lambda_k j} \ket{j}\ket{\psi_k}$.
Next, an inverse $QFT$ operation is performed on the ancilla register.  The result left on the ancilla register are estimates of the 
eigenvalue parameter, $\lambda_k$, specifically $\ell_k = 2^m \lambda_k$, in superposition (as dictated by the input vector $\ket{\Psi}$).  For a more 
detailed description of the eigenvalue estimation algorithm see~\cite{KLM}.

For stage two, we apply the controlled operation $c-\mathcal{R}(2 \pi \ell_k t/2^m)$ to induce the phase shift $e^{2 \pi i \ell_k t/2^m}$ when the eigenvalue parameter estimate $\ell_k$ is in the control register.  This computation corresponds to
the exponentiation of the diagonal matrix in the spectral decomposition formula.  There are a few natural ways to implement this step, as shown in
\cite{CEMM98} or \cite{zalka}.

Finally, the state of the control register (containing the $\ell_k$ values) is uncomputed back to the all-zeros (input ancilla) state.  This completes the third and final stage of the computation.

This leaves the registers in the final state $\ket{\mathbf{0}}\otimes\sum_k \alpha_k e^{2\pi i\lambda_k t} \ket{\psi_k}$, which is simply the
result of the application of $\UU^t$ to the state $\ket{\Psi}$.  When we drop the assumption that the eigenvalue parameters
$\lambda_k$ are exact ratios of some integers $\ell_k$ divided by $2^m$, we will in general get precision errors in the eigenvalue estimation.  This imprecision in 
the resolution of $\lambda_k$ means the phases applied to the state are not exact, which further implies that the uncomputation 
step does not return the ancilla register precisely to all zeroes.
 Still, by a proper choice of parameters, and a ``gap''
assumption discussed below, the errors can be managed---in particular, we prove that they are exponentially small in the parameter $m$.
 This error reduction is done by choosing an eigenvalue estimation algorithm that outputs an estimate with error at most $\frac{1}{2^m}$ with probability 
in $1 - O(\frac{1}{2^{m}})$. This can be done with $O(m)$ repetitions of the standard QFT-based eigenvalue estimation algorithm \cite{KLM} and applying 
the Chernoff bound. This uses  $O(m 2^m)$ calls to the black box.
Relevant calculations may be found in \ref{sec:complerrors}.

\subsection{Underlying Assumptions}

Our algorithm, as formulated in the previous section, asserts that in addition to the black box implementation of $\UU$ itself we also have access to a 
black box implementing controlled-$\UU$ and a black box implementing $\UU^{-1}$. Furthermore, we also must assume, as discussed in more detail in the 
following section, that spectrum of matrix $\UU$ has a small gap, in particular, that $\UU$ has no eigenvalue $\lambda=e^{i\phi}$, where $\phi \in
(2\pi(1-g),\;2\pi)$ for some value $g$.  For simplicity, we assume that $g \geq \frac{1}{2^m}$; otherwise, we would need to replace some of the $O(2^m)$ 
terms with $O(\max \{2^m, \frac{1}{g} \})$ (the reason is that we need to do eigenvalue estimation with precision less than the gap size with high probability).   In 
this case, we achieve an approximation with error in $O(\frac{1}{2^m})$.  Note that there are no other assumptions about gaps elsewhere in the spectrum. Furthermore, an 
assumption of this form is essentially necessary in the case of worst-case complexity (as we
explain in Section \ref{optimality}), which is what we are interested in.  Average-case performance is discussed more in \ref{app:disc}. 
 
Not all of the above restrictions are necessary, but they facilitate a clear and transparent description of the algorithm and its analysis.  In some cases of interest, the individual 
restrictions may be dropped.  In particular, in cases where only the black box implementation 
of $\UU$ is given (and not the controlled-$\UU$), one can exploit the technique for implementing a controlled-$e^{-i \phi} U$, where $e^{i \phi}$ is a 
random eigenvalue of $\UU$, and use it in the algorithm to find arbitrary powers, as outlined in \cite{Kit95} (see \ref{app:cu} for details).  In 
some special cases, we can remove the assumption of having a black box for $\UU^{-1}$. For example, for non-integer values of $t \geq 2^m$, we can approximate 
$\UU^{t}$ with error in $O(\frac{1}{t})$ using $O(t)$ calls to the controlled-$\UU$, without any use of a black box for $\UU^{-1}$ (see \ref{app:inv}).
We note that the above restriction on the form of the spectrum is
naturally satisfied in many practical unitaries, including the quantum Fourier transform, standard oracles $\ket{j} \mapsto (-1)^{f(j)} \ket{j}$ for 
computing Boolean functions $f(\cdot)$, and oracles such as those implicitly used in the adiabatic algorithms of Farhi {\it et al.} \cite{FGG} that 
reversibly compute a function with a finite discrete spectrum.

It is also worth noting that, for non-integer values of the power $t$, there are many operators that one might reasonably call the $t^\textrm{th}$ power of 
$\UU$. In other words, for any given $\UU$, there are many Hamiltonians $H$ such that $\UU = e^{-i H}$, and thus one could naturally define $\UU^{t}$ as 
$e^{-i H t}$ for
any one of these Hamiltonians.  For instance, consider $t=\frac{1}{2}$ and the identity matrix $\mathbb{I}=\left(1\;\;0\atop 0\;\;1\right)$.  Let us find 
all unitary matrices $B$ such that $BB=\mathbb{I}$.  In the most natural understanding, each such $B$ is a possible square root of $\mathbb{I}$.  Firstly, 
matrices $\mathbb{I}$ and $\left(-1\;\;\;0\atop 0\;\;\;-1\right)$ are both square roots of $\mathbb{I}$. This happens because there are two possible 
square roots of the complex number $1$, which is the eigenvalue of $\mathbb{I}$.  In addition, each eigenspace of a unitary matrix may be broken into 
subspaces such that different square roots of the corresponding eigenvalue may be used to construct square roots of a given matrix.  In our case, this 
helps to construct two more square roots of the form $\pm \left(1\;\;\;0\atop 0\;-1\right)$.  Furthermore, every matrix of the form $\pm \left(\cos 
a\;\;\;\;\;\; e^{b}\sin a \atop e^{-b}\sin a \;\; -\cos a\right)$, and its transpose, where $a$ and $b$ are real valued parameters is a square root of the 
operator $\mathbb{I}$.  For the choice of parameters $a=\frac{\pi}{2}$ and $b=0$ this allows to construct root $\left(0\;\;1\atop 1\;\;0\right)$, also 
known as the Pauli-X matrix.  Roots of the identity matrix in higher dimensions get more complicated.  Thus, the problem of finding fractional powers is not 
completely straightforward to define.  In our solution, we construct fractional powers based on the spectral 
decomposition, choosing the primitive complex fractional power of every eigenvalue (though, in principle our method allows us to break the eigenspaces into subspaces and take different roots of the eigenvalue for each subspace).  Thus, in a sense, the 
fractional power we construct is the most natural one, and also, it is the fractional power of a matrix most commonly defined in
linear algebra textbooks. With the above example, our algorithm 
implements the $\mathbb{I}$ root of $\mathbb{I}$, though modifications are possible that would allow us to explore a wider range of square roots of 
$\mathbb{I}$.

\subsection{Optimality} \label{optimality}

For such a general black box and assuming a spectral gap of size $g = \frac{1}{2^m}$, we cannot compute $\UU^{\frac{1}{2}}$ with constant error with fewer 
than $\Omega(2^m)$ calls to $\UU$.  To see why this is the case, suppose we are given either $\UU = \mathbb{I}$ or $\UU = \UU_{1 - 1/2^{m}} = 
\ket{0}\bra{0} + e^{- 2\pi i/2^m} \ket{1}\bra{1}$.  It follows from the following simple lemma that  $\Omega(2^m)$ calls to $\UU$ are necessary to 
correctly guess which $\UU$ we were
given with probability greater than $\frac{2}{3}$.

{\bf Lemma 1:} {\it Suppose we are given a black box $\UU$ that implements either $\mathbb{I}$ or
$\UU_{1-\delta} = \ket{0}\bra{0} + e^{-i \delta} \ket{1}\bra{1}$, for some small $\delta > 0$. Any algorithm that correctly guesses the identity of $\UU$ 
with probability at least $\frac{2}{3}$, for any prior distribution of the two possible values of $\UU$, must make $\Omega(\frac{1}{\delta})$ evaluations 
of $\UU$.}

This can be easily proved, {\it e.g.}, by the ``hybrid'' method \cite{BBBV}.

Note that the root $\UU_{1-1 / 2^m}^{\frac{1}{2}} (\ket{0}+\ket{1}) = \ket{0} - e^{- \pi
i/2^m}\ket{1}$ is almost perfectly distinguishable from $\mathbb{I}^{\frac{1}{2}} (\ket{0} + \ket{1}) = \ket{0} + \ket{1}$.  Thus, the power finding 
algorithm must take
$\Omega(2^m)$ calls to $\UU$ in order to compute $\UU^{\frac{1}{2}}$ with (small) constant error.  The same holds for any power $t \in (0,1)$ that is
bounded away from from both $0$ and $1$.

Thus, while more efficient ways to compute fractional powers of $\UU$ are possible in special cases, in general one cannot do much better than what we
have described.
In the case when $\frac{1}{2^m} < g < \pi$, we can still obtain a lower bound of $\Omega(2^m)$ queries by letting $\UU$ be either $Z$ or $\ket{0}\bra{0} -
e^{-2 \pi i/2^m} \ket{1}\bra{1}$, and computing square roots of $\UU^2$ as a means of distinguishing the two cases.

This same argument also shows why a gap assumption is necessary.
Without the gap assumption, there is no upper bound on the worst-case complexity of computing the square root of a black box unitary, since such a 
black box could be used to distinguish $\mathbb{I}$ from
$\UU_{1-\delta} = \ket{0}\bra{0} + e^{-i \delta} \ket{1}\bra{1}$ for arbitarily small values of $\delta$.
As pointed out earlier, if we have a known gap $g$ without a promise that $g \in \Omega(\epsilon) =
\Omega(1/2^m)$, then a complexity of $O(\frac{1}{\epsilon} \log \frac{1}{\epsilon})$ suffices,
and the example above shows that $\Omega(\frac{1}{g})$ queries are
necessary.  If $g$ is unknown then there are several cases one might consider.  If we are promised a lower bound $0 < g_{\mathrm{min}} \leq g$, then in the worst case, we must use $O(\frac{1}{g_{\mathrm{min}}} \log \frac{1}{\epsilon})$ calls to the unitary.  If instead we have an means of reliably
recognizing when the algorithm has succeeded, then trying a sequence of lower bounds that decreases exponentially until the algorithm succeeds will also give a complexity in $O(\frac{1}{\min \{g,
\epsilon\}} \log \frac{1}{\epsilon})$, which is the same complexity as when $g$ is known.  If, however, we do not know $g$ and have no means of recognizing when the algorithm has succeeded, then no upper bound is possible.

\section{Large Powers and an Example of Exponential Improvement} \label{sec:example}

When it comes to computing large powers of $\UU$ by this method, precision in the eigenvalue estimates becomes a limiting factor.  If instead of the exact 
eigenvalue $e^{2 \pi i \lambda_k}$ we have its estimate $e^{2 \pi i \widetilde{\lambda_k}}$, where $|\widetilde{\lambda_k} - \lambda_k| = \epsilon$, then 
mapping $\ket{\psi_k} \mapsto e^{2 \pi i \widetilde{\lambda_k} t}\ket{\psi_k}$ will lead to an error in the phase parameter of size $t \epsilon$.  For arbitrary 
$\UU$, it takes $O(\frac{1}{\epsilon})$ calls to $\UU$ in order to get an estimate with error at most $\epsilon$ with high probability.  Thus to achieve an error of $\delta = t \epsilon$ in the application of $\UU^t$, we need $\epsilon  = \delta/t$ precision in the 
estimate, amounting to $O(\frac{t}{\delta})$ calls to $\UU$.  For $t > 1$, a better thing to do in this general case is to first apply $\UU$ a total of $\lfloor t 
\rfloor$ times, and then use the method from the previous section in order to compute the remaining fractional power of $\UU$.

However, if for special cases of $\UU$ we are able to get smaller errors in the eigenvalue estimates, we can do much better. For example, for black boxes 
that map $\ket{j} \mapsto (-1)^{X_j} \ket{j}$ for some $j \in \{0,1\}^n$ and $X_1 X_2 \ldots  X_j \ldots X_N \in \{0,1\}^N$, it is easy to exactly 
determine the eigenvalue parameter $X_j$ with only one call to the black box.  As discussed later, it is easy to exactly determine the eigenvalue of an 
eigenvector of the quantum Fourier transform, since the QFT has order four. In both of these examples, our technique does not help with computing higher 
powers of $\UU$ since $\UU$ has very low order, say $r$, and thus computing $\UU^{t}$ is equivalent to computing $\UU^{t \mod r}$, where $t \mod r$ is the 
unique element of $t - r\mathbb{Z}$ that lies in in the interval $[0,r)$.  We next construct an example where $\UU$ has very high order, but one can still 
compute precise estimates of its eigenvalues with relatively few calls to $\UU$, and thus exponentiate $\UU$ with much fewer than $t$ calls, even for $t$ 
less than the order of $\UU$.

Suppose the eigenvalues of the desired unitary $\UU$ are all of the form $e^{2 \pi i \frac{\ell_k}{p_k}}$, for $k \in \{1, 2, \ldots, b\}$, where $\ell_k$ 
is an integer, $p_k \in \{p_1=2,p_2=3,p_3=5,...,p_b | p_i=i^{\textrm{th}}\textrm{ prime}\}$, and $b$ is a natural number.  Then $\UU^B = \mathbb{I}$, where 
$B=\prod_{k=1}^{b} p_k$, and (assuming that each $p_k$ occurs at least once with a non-zero $l_k$) no lower power of $\UU$ is equal to the identity.  If 
$t$ is on the order of $B$, then the number of calls to the unitary required for the straightforward implementation of $\UU^t$ is in $O(B) = 
O(\prod_{k=1}^{b} k\log k)$.  This product is called a {\it primorial}~\cite{prim}, and it is in $O(e^{(1+o(1))b \log b}) = O(b^{(1+o(1))b})$.

We next demonstrate how our basic algorithm introduced in Section~\ref{sec:algo} may be updated to approximate $\UU^t$ with exponentially fewer calls 
to $\UU$ than the number of calls it takes to obtain $\UU^t$ via a direct application of $t$ unitaries.

The method used for this problem employs the continued fraction algorithm, which was also used in the order-finding part of Shor's algorithm~\cite{shor}.  The eigenvalue estimation algorithm will return an estimate $\frac{h}{2^m}$ of the eigenvalue parameter 
$\frac{\ell_k}{p_k}$ to within an error at most $\frac{1}{2^m}$ with probability at least $\frac{8}{\pi^2}$.  The error probability can be amplified down 
to $\frac{1}{2^m}$ by repeating the estimation $O(m)$ times and taking the majority vote.  If $m$ is chosen such that $m \in O(\log b)$, 
so that $2^m > 2 p_b p_{b-1}$, then the fraction $\frac{\ell_k}{p_k}$ will be the only fraction, or {\it convergent}, in the continued fraction expansion 
of $\frac{h}{2^m}$ that has distance at most $\frac{1}{2^m}$ from $\frac{h}{2^m}$ and has a denominator at most $p_{b}$.

\begin{figure}[h!]
\begin{center}
\includegraphics[width=3in]{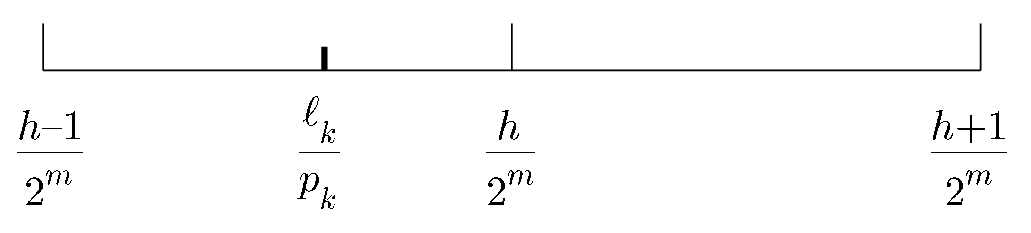}
\caption{A diagram illustrating the procedure.  The estimate is $\frac{h}{2^m}$.  The smallest gap between any two eigenvalues is $\frac{1}{p_b p_{b-1}}$and we require this to be larger than $\frac{2}{2^m}$ to ensure that only one rational fraction with denominator no greater than $p_b$ will be found in the region $\frac{h-1}{2^m}$ to $\frac{h+1}{2^m}$.}
\label{contfrac}
\end{center}
\end{figure}

This convergent $\frac{\ell_k}{p_k}$ can be found with $O(m)$ arithmetic operations over integer numbers with at most $\log m$ bits each ({\it e.g.}, no more than $O(m^3)$ binary operations using trivial algorithms for division and multiplication) using the continued fraction algorithm.  Once we have the 
exact value of the eigenvalue parameter $\frac{\ell_k}{p_k}$, with probability $\frac{1}{2^{m}}$, we can correctly compute {\it any} positive power $t$ of 
$\UU$ with error in $O(\frac{1}{2^{m}})$ using
$poly(b, \log t)$ calls to $\UU$ and other elementary operations.  This is done by mapping, in superposition, $\ket{\ell_k}\ket{p_k} \ket{\psi_k} \mapsto e^{2 \pi 
i t \frac{\ell_k}{p_k} + O(\frac{1}{2^m})} \ket{\ell_k}\ket{p_k} \ket{\psi_k}$, which can be done by standard reversible computing methods and $m$-bit 
arithmetic operations (the $O(\frac{1}{2^m})$ error is due to precision in the arithmetic operations). Then, as usual, one uncomputes the computation of $\ket{\ell_k}\ket{p_k}$.  As alluded to before, this gives an overall complexity 
of $O(m 2^m) = O(b \log b)$ to perform the operation up to error $O(\frac{1}{2^{m}})$.  This follows from~(\ref{eq:error}) in~\ref{sec:complerrors}.

This complexity is polynomial in $\log t$ (vs $poly(t)$) because we only need to compute an $m$-bit approximation to $t \ell_k/p_k$.   This also allows us to uncompute the eigenvalue estimates found in the first 
stage of the algorithm without ever needing to use the operation $\UU^{-1}$.  Note that $t$ could be very large, for example $t \sim B/2 = 2^{b-1}$.  In this case, we can apply $\UU^t$ to any state of the appropriate dimension with the output state containing an error term with probability amplitude $O(\frac{1}{b})$ with only $O(b\log b) <\!\!\!< t \in O(2^{b})$ calls. The error probability can efficiently be made exponentially small by repeating the phase estimation part.

\section{Fractional Quantum Fourier Transform}
\label{sec:ft}

In classical computation, roots of the Fourier transform are used for efficient filtering of the frequency noise that is some non-constant function of a conjugate variable, such as time \cite{ar:a}.  Given the role the QFT plays in the quantum computation, and the importance of roots of Fourier transforms in classical computation, there may as well be 
other interesting techniques based on the fractional powers of the quantum Fourier transform.

Note that $QFT^4 = \mathbb{I}$.  Let us illustrate the computation of a fractional power $t$, $t<1$ of the QFT using 
the algorithm introduced in Section \ref{sec:algo}.  First, note that the QFT has four distinct eigenvalues: $+1,\;i,\;-1$ and $-i$.  This means that we 
need only two ancilla bits for the eigenvalue estimation part of the exponentiation algorithm, and such estimation results in the exact values.  During 
the eigenvalue estimation algorithm, the controlled-QFT is applied three times ($2^0=1$ with the first ancilla qubit as a control, and $2^1=2$ more with 
the second ancilla as a control).  We need three more applications of the controlled-QFT in the third stage of the algorithm to uncompute ancilla. ÊThis means that the query complexity of the computation of a fractional power of the QFT is exactly six queries to the controlled-QFT.  Moreover, during such computation we 
introduced no errors due to the approximations of eigenvalues, and thus, given all gates are perfect, the computed fractional power is exact.

This specific example of fractional QFT was alternatively derived as a special case of the work of Klappenecker and R\"otteler~\cite{kr}, where they address the different problem of how to simulate a 
known unitary matrix $A$ given other operators that generate an algebra containing $A$.

\section{Conclusions and Discussion}

In this work we have further developed the connection between continuous time black boxes ({\it i.e.}, ``Hamiltonian oracles'') and their discrete counterparts, unitary black boxes.  Previous work has mostly focused on important special cases of this relationship.  Our work pushes the boundary in a new direction, where the unknown $\UU$ is a unitary acting on an arbitrarily large Hilbert space.  It also demonstrates constructively that an algorithm with a continuous oracle can also be implemented with a discrete oracle that implements the continuous oracle for some fixed time (with some assumptions
on the Hamiltonian encoded by the oracle; {\it e.g.}, if the energies, the eigenvalues of the Hamiltonian, are between 0 and $E_{max}$, then the fixed time interval should be bounded below $2 \pi / E_{max}$ so that our gap assumption is satisfied).

For the case that $t$ is a positive non-integer, the complexity of this algorithm is substantially higher than that in the continuous setting, where we have access to a Hamiltonian $H$ satisfying $\UU = e^{-i H}$.  If we have direct access to the Hamiltonian, $H$, then in principle we can easily perform 
any unitary of the form $e^{-i Ht}$ with a cost $t$ in terms of calls to the Hamiltonian.  For an algorithm with total query complexity $t < 1$ in the Hamiltonian black box model, it is hard to imagine a reasonable way to effect $e^{-i H t}$ with fewer than one evaluation of $\UU$, and indeed, our algorithm makes $O(\frac{1}{\epsilon} \log \frac{1}{\epsilon})$ calls to $\UU$ to effect $e^{-i Ht}$ with precision $\epsilon$ (we show a corresponding lower bound in section \ref{optimality}), which is 
notably lower than the number of calls required in the obvious approach of doing process tomography on $\UU$, approximating $\UU^{t}$, and then synthesizing a circuit to implement such an approximation. In particular, our approach does not depend on the dimension of the Hilbert space that $\UU$ acts on.

We have presented an algorithm for finding the powers of unknown unitary operations raised to any real power $t > 0$.  We found the complexity of the algorithm to be in $O(\lfloor t \rfloor + 2^m m)$ where the error is in $O(\frac{1}{2^m})$.  For large integers $t$ we presented a non-trivial example where our method is exponentially more efficient than the direct repeated application of $\UU$ a total of $t$ times.  Note that the same exponential speed-up is possible also in the continuous case if the Hamiltonian has some structure that enables eigenvalue estimation much better than the standard worst-case limits; in such cases, our procedure, using a modest amount of ancilla workspace, can simulate the evolution of $H$ for a time $\tau$ in time polynomial in $\log (\tau)$.

Our algorithm could be of practical use in a situation where we have very precise clocks and control, say with error less than a very small positive $\epsilon$, but for some technical reason we must run the unknown Hamiltonian for a minimum time $T_{min} >> \epsilon$ ({\it e.g.}, because of a
minimum reset time on some critical piece of apparatus, like a detector, used to simulate or execute the Hamiltonian), and $T_{min}$ is still smaller than $2 \pi / E_{max}$ (assuming the energies are non-negative).  In this case, there is a smallest amount of time for which a given Hamiltonian can naturally be run.  If we wish to run the Hamiltonian for less than that amount of time, then our method could be used.

Further, if the Hamiltonian that solves some problem is time dependent, as is often the case in quantum circuits or in models based on the permutation or braiding of particles on a surface, it may be of interest to apply some fraction of this total transformation to turn the solution to the problem into a subroutine of a larger algorithm.  The method we propose provides a general  way of obtaining this fractional application.

Our technique may also be applied to compute other functions of $\UU$, by replacing each eigenvalue $e^{i \lambda_k}$ with $e^{i f(\lambda_k)}$ for any function $f(\cdot)$.  This may be useful in contexts such as Childs' method~\cite{childs} for simulating continuous time quantum walk Hamiltonians with discrete time walks.  This requires applying phases that are the sines or arcsines of the eigenvalues of the original Hamiltonian.  Our method could achieve this for an unknown Hamiltonian to be simulated.  The properties of the function $f(\cdot)$ will determine the efficiency as a function of the precision.  In the case when $f(\cdot)$ is continuous with bounded first derivative, and $2\pi$-periodic, then we can also drop the assumption that there must be a gap in the spectrum.

In addition, there are a number of classical cryptography schemes that rely on the computational difficulty of finding roots modulo some number, such as RSA.  One might naturally attempt to construct quantum cryptography schemes where knowledge of the ``secret key'' corresponds to the ability to compute a fractional power ({\it i.e.}, a root) of some unitary operator. This algorithm provides a way of implementing such roots, implying serious limitations for such an approach in quantum cryptographic schemes.

\section*{Acknowledgements}

We thank Scott Aaronson, Patrick Hayden, Rolando Somma, and John Watrous for discussions about the relevance of this problem. Dr. Aaronson, Dr. Hayden, and Dr. Somma were interested in black box square root question, and Dr. Watrous sketched a solution similar to the one that we had.  We thank Dr. Donny Cheung at the University of Calgary for helpful insights.    This research has been supported by DTO-ARO, NSERC, CFI, CIFAR, CRC, ORF, Ontario-MRI, QuantumWorks, and the Government of Canada.  L. Sheridan gratefully acknowledges the support of the Mike and Ophelia Lazaridis Fellowship.

This article was based on work partially supported by the National Science Foundation, during D. Maslov's assignment at the Foundation. Any opinion, finding and conclusions or recommendations expressed in this material are those of the author and do not necessarily reflect the views of the National Science Foundation.

\appendix

\section{Controlled Unitaries}\label{app:cu}

Note that the eigenvalue estimation algorithm requires the use of a controlled unitary $c-\UU$.  If we are given an eigenvector of $\UU$ with eigenvalue 
$+1$, then this transformation can be implemented using a
controlled-SWAP (equivalently, a series of Fredkin gates \cite{FT}), as was shown by Kitaev \cite{Kit95} and is illustrated in the following circuit 
diagram:

\[
\Qcircuit @C=1.5em @R=2em {
\lstick{\ket{q_1}_{\textrm{control}}} & \qw & \ctrl{1} & \qw & \ctrl{1} & \qw   \\
\lstick{\ket{q_2}_{\textrm{target}}} & \qw & \multigate{1}{SWAP} & \qw & \multigate{1}{SWAP} & \qw \\
\lstick{\ket{\psi_k}} & \qw & \ghost{SWAP} & \gate{\UU} & \ghost{SWAP} & \qw \\
}
\]

It is easy to verify that the above circuit implements
\[ \ket{0}_{\textrm{control}} \ket{q_2}_{\textrm{target}} \ket{\psi_k}
\mapsto e^{2 \pi i \lambda_k}  \ket{0}_{\textrm{control}} \ket{q_2}_{\textrm{target}} \ket{\psi_k} \]
\[ \ket{1}_{\textrm{control}} \ket{q_2}_{\textrm{target}} \ket{\psi_k}
\mapsto \ket{1}_{\textrm{control}} (\UU \ket{q_2}_{\textrm{target}}) \ket{\psi_k}  . \]

In other words, up to a global phase, this implements the controlled-$(e^{-2 \pi i \lambda_k}\UU)$. If $\lambda_k = 0$, {\it i.e.}, $\ket{\psi_k}$ has 
eigenvalue $+1$, then we have implemented the $c-\UU$. Otherwise, we have a very similar operation, which may or may not be sufficient, depending on the 
application.
If, for example, any $c-(e^{i \phi} \UU)$ will suffice, as long as we use the same one consistently, then this is easy to achieve.

We do not need to assume that we know the eigenvectors $\{ \ket{\psi_k} \}$ of $\UU$ or are given a specific eigenvector.   Instead, we can use any state in the target register. For simplicity, we can consider the completely mixed state, which can be decomposed as an equal mixture over 
the eigenstates.  Provided we keep the state that is in the target register, and re-use the resulting state every time we construct a controlled-$\UU$ 
between these two registers, the phase (though random) will be the same for
the entire computation.  That is, this is equivalent to implementing a $c-(e^{i \phi}\UU)$ consistently throughout the computation, where $\phi$ is an 
unknown random value (whose distribution depends on the weight of the eigenspace of $e^{-i \phi}$ in the initial mixed state). In essence, the target 
system serves as a phase reference.

\section{Complexity and Error Analysis} \label{sec:complerrors}

We will describe upper bounds on the precision of our algorithm, which is constrained by the lack of perfect precision in computing the eigenvalue estimates. We 
give the analysis for $t= \frac{1}{2}$, however the same final bound applies for any $t \in (0,1)$.  

Let us first describe an ``ideal'' algorithm that exactly computes the square root of $\UU$. For the ideal algorithm, some of the steps will be 
unnecessary, but we include them to make the comparison with the actual algorithm simple.

Stage 1 computes a near-optimal eigenvalue estimation (in the sense of nearly optimizing the chance of obtaining an estimate with error at most
$\frac{1}{2^{2m}}$).
We do this by repeating the ``standard'' eigenvalue estimation algorithm \cite{CEMM98} a total of $r \in O(m)$ times and take the majority answer, with
the constant chosen so that each eigenvalue estimate $e^{2 \pi i \widetilde{\lambda_k}}$ of $e^{2 \pi i \lambda_k}$ satisfies $|e^{2 \pi i \widetilde{\lambda_k}}-e^{2 \pi i 
\lambda_k}| \leq \frac{1}{2^m}$ with probability in $1 - O(\frac{1}{2^{2m}})$.
One can fine-tune the optimal eigenvalue estimation procedure further ({\it e.g.}, \cite{DDEMM07}), however, this simple procedure is sufficient.

This eigenvalue estimation procedure is described by the transformation
\begin{equation}
\ket{0} \sum_k \alpha_k \ket{\psi_k} \mapsto \sum_k \alpha_k \left( \sum_{\bf{y}} \beta_{{\bf{y}},k} \ket{\bf{y}} \ket{\lambda_{{\bf{y}}}}  \right)
\ket{\psi_k}
\end{equation}
where the values ${\bf{y}} \in \mathbb{Z}_{2^m}^r$ (where $\mathbb{Z}_{2^m} =  \{0,1,\ldots, 2^m-1\}$) are $r$-tuples of integers, and $\lambda_{\bf{y}}$ 
is the value obtained by taking the most commonly occurring integer $y_{\textrm{mode}}$ in the $r$-tuple of integers ${\bf{y}}$ (with some convention for 
breaking ties) and letting
$\lambda_{\bf{y}} =  y_{\textrm{mode}}/2^m$.

In the ideal Stage II, we exactly implement the map
\begin{eqnarray}
\sum_k \alpha_k \left( \sum_{\bf{y}} \beta_{{\bf{y}},k} \ket{\bf{y}}  \ket{\lambda_{{\bf{y}}}} \right)  \ket{\psi_k}  \nonumber  \\ 
\ \ \ \ \ \ \ \ \ \ \ \ \ \ \mapsto \sum_k \alpha_k \left(
\sum_{\bf{y}} \beta_{{\bf{y}},k} \ket{\bf{y}}  \ket{\lambda_{{\bf{y}}}} \right)  e^{2 \pi i \lambda_k/2} \ket{\psi_k}  .
\end{eqnarray}

Then Stage III uncomputes the eigenvalue estimation exactly (since with the ideal Stage II, there is no coupling of the control registers with the target 
registers), leaving us with the ideal output
\begin{equation}
\sum_k \alpha_k  \ket{00 \ldots 0} e^{2 \pi i \lambda_k/2} \ket{\psi_k}  .
\end{equation}

The only difference between this ideal algorithm and our actual algorithm is that Stage II actually effects the unitary
\begin{equation}
\sum_k \alpha_k \left( \sum_{\bf{y}} \beta_{{\bf{y}},k} \ket{\bf{y}}  \ket{\lambda_{{\bf{y}}}} \right)  \ket{\psi_k} \mapsto  \sum_k \alpha_k  \sum_{\bf{y}} 
\beta_{{\bf{y}},k} \ket{\bf{y}}  \ket{\lambda_{{\bf{y}}}}  e^{2 \pi i \lambda_{{\bf{y}}}/2} \ket{\psi_k}  .
\end{equation}

Since the probability that  $|e^{2 \pi i \lambda_{\bf y}/2} - e^{2 \pi i \lambda_k/2}| > \frac{1}{2^m}$ is in $O(\frac{1}{2^m})$ (note that here we use the ``gap'' assumption), we can write the state in Stage II as
\begin{eqnarray}
\ket{\Phi} &=&  \sum_k \alpha_k  \sum_{\bf{y} \in S} \beta_{{\bf{y}},k} \ket{\bf{y}}  \ket{\lambda_{{\bf{y}}}}  e^{2 \pi i \lambda_{{\bf{y}}}/2} \ket{\psi_k}     \nonumber  \\
& & \ \ \ \ \ + \sum_k \alpha_k \sum_{\bf{y} \in S'} \beta_{{\bf{y}},k} \ket{\bf{y}}  \ket{\lambda_{{\bf{y}}}}   e^{2 \pi i \lambda_{{\bf{y}}}/2} \ket{\psi_k} ,
\end{eqnarray}
where the values of $\bf{y} \in S$ include the ``good'' values of $\bf{y}$ that produce estimates $\lambda_{{\bf{y}}}$ satisfying $|e^{2 \pi i 
\lambda_{{\bf{y}}}} - e^{2 \pi i \lambda_k}| \leq \frac{1}{2^m}$, and the ``bad'' values, $\bf{y} \in S'$, are the rest.
The norm of the bad part of the state is in $O(\frac{1}{2^{2m}})$ and the norm of the ``good'' part is $1-O(\frac{1}{2^{2m}})$.

Let $\Lambda$ denote the ideal phase shift operator, and $\tilde{\Lambda}$ denote the actual operator.

We thus have
\begin{eqnarray*}
\Lambda \ket{\Phi} - \tilde{\Lambda} \ket{\Phi} &=& \sum_k \alpha_k
\sum_{\bf{y} \in S} \beta_{{\bf{y}},k} \ket{\bf{y}} \ket{\lambda_{{\bf{y}}}}  (e^{2 \pi i \lambda_k/2} - e^{2 \pi i \lambda_{{\bf{y}}}/2}) \ket{\psi_k}  \\
& & \ \ + \sum_k 
\alpha_k \sum_{\bf{y} \in S'} \beta_{{\bf{y}},k} \ket{\bf{y}}  \ket{\lambda_{{\bf{y}}}}   (e^{2 \pi i \lambda_k/2} - e^{2 \pi i \lambda_{{\bf{y}}}/2}) \ket{\psi_k} . \nonumber
\end{eqnarray*}
Thus we can bound the norm of this difference by
\begin{equation}
|| \Lambda \ket{\Phi} - \tilde{\Lambda} \ket{\Phi} ||^2  \leq \sum_k
\sum_{\bf{y}\in S} |\alpha_k \beta_{{\bf{y}},k}|^2  |\delta_{k, {\bf{y}}} |^2 + \sum_k \sum_{\bf{y} \in S'} |\alpha_k \beta_{{\bf{y}},k}|^2  |\delta_{k,
{\bf{y}}} |^2 ,
\end{equation}
where  $|\delta_{k, {\bf{y}}} |^2 = |e^{i \lambda_k/2} - e^{i \lambda_{{\bf{y}}}/2}|^2$.

We know that $\sum_{\bf{y} \in S} \sum_k |\alpha_k \beta_{{\bf{y}},k}|^2 \in 1 - O(\frac{1}{2^{2m}})$, $\sum_{\bf{y} \in S'} \sum_k |\alpha_k
\beta_{{\bf{y}},k}|^2 \in O(\frac{1}{2^{2m}})$, and that  $|\delta_{k, {\bf{y}}} |^2 \in O(\frac{1}{2^{2m}})$ for the good values, and at most $1$ for the
bad values.

Thus
\begin{equation} || \Lambda \ket{\Phi} - \tilde{\Lambda} \ket{\Phi}||^2  \in O\left(\frac{1}{2^{2m}}\right).
\end{equation}

Noticing that
\begin{equation}
||\ket{u} -\ket{v}||^{2}_2 = (\bra{u}-\bra{v})(\ket{u}-\ket{v}) \geq 2(1-|\bracket{u}{v}|)
\end{equation}
and using the equality
\begin{equation}
||\ket{u}\bra{u} -\ket{v}\bra{v}||_{\Tr} = 2\sqrt{1-|\bracket{u}{v}|^2} ,
\end{equation}
which implies that
\begin{equation}
||\ket{u}\bra{u} -\ket{v}\bra{v}||_{\Tr} \leq 2 ||\ket{u} -\ket{v}||_2 ,
\end{equation}
so $|| \Lambda(\ket{\Phi}\bra{\Phi})\Lambda - \tilde{\Lambda}(\ket{\Phi}\bra{\Phi})\tilde{\Lambda}||_{\Tr}  \in O(\frac{1}{2^{m}})$.  The trace norm of the difference between the two unitary superoperators is simply given by taking the maximization over states $\ket{\Phi}$.  Since this holds for any state $\ket{\Phi}$, we have
  \begin{equation} \label{eq:error} || \Lambda - \tilde{\Lambda}||_{\Tr}  \in O\left(\frac{1}{2^{m}}\right). \end{equation}
This is the case after Stage II of the algorithm.  Note that Stage III is a unitary operation that is the 
same for both the ideal and the actual case.  The trace norm is invariant under unitary transformations, so the trace norm of the unitary operators for
the full ideal and actual algorithms have the same bound.  In the case of unitary operators this is equivalent to the diamond norm.

\section{Inverses} \label{app:inv}

Our algorithm for generating $\UU^{t}$ uses $\UU^{-1}$ when
the eigenvalue estimates need to be uncomputed.
More specifically, we start with $\ket{0} \ket{\psi_k}$, and we apply $(QFT^{-1}_{2^m} \otimes \mathbb{I}) c-\UU^j (QFT_{2^m} \otimes \mathbb{I})$ to 
compute
\begin{equation}
\sum_j \left(QFT_{2^m} \frac{e^{2 \pi i j \lambda_k}}{\sqrt{2^m}}  \ket{j}\right) \ket{\psi_k} .
\end{equation}
 We then use the value in the first register to control the phase shift corresponding to $\ket{\psi_k}$.  For simplicity, let us assume that we do not  
bother to induce a phase shift, but we are still left with the task of uncomputing the ``damage'' we have done. It is trivial if we have access to 
$c-\UU^{-1}$ (which can be implemented with a $c-\UU$ and $\UU^{-1}$), since we just apply
 $(QFT^{-1}_{2^m} \otimes \mathbb{I}) c-\UU^{-j} (QFT_{2^m} \otimes \mathbb{I})$ to get back $\ket{0} \ket{\psi_k}$.

However, in some cases, we do not need the $\UU^{-1}$ operations.
We could attempt to circumvent the need for $\UU^{-1}$ operations by applying
 $(QFT_{2^m} \otimes \mathbb{I}) c-\UU^{j} (QFT^{-1}_{2^m} \otimes \mathbb{I})$ to
 yield
 \begin{eqnarray} 
 (QFT_{2^m} \otimes \mathbb{I}) c-\UU^{j} (QFT^{-1}_{2^m} \otimes \mathbb{I})(QFT^{-1}_{2^m} \otimes \mathbb{I}) c-\UU^j (QFT_{2^m} 
\otimes \mathbb{I}) \ket{0} \ket{\psi_k} \nonumber \\
 = (QFT_{2^m} \otimes \mathbb{I}) c-\UU^{j} (M \otimes \mathbb{I}) c-\UU^j (QFT_{2^m} \otimes \mathbb{I}) \ket{0} \ket{\psi_k} 
 \end{eqnarray}
 where $M = QFT^{-2}_{2^m}$, which implies $M\ket{x} = \ket{2^m - x \mbox{ mod } 2^m}$.

We can easily verify that
$(M \otimes \mathbb{I})c-\UU^{j} (M \otimes \mathbb{I}) = c-\UU^{2^m - j \mbox{ mod } 2^m}$, and thus the above state equals
 \begin{eqnarray}
= (QFT_{2^m} \otimes \mathbb{I}) (M \otimes \mathbb{I}) c-\UU^{2^m - j \mbox{ mod } 2^m} ( c-\UU^j (QFT_{2^m} \otimes \mathbb{I}) \ket{0} \ket{\psi_k}  
\nonumber \\
  = (QFT^{-1}_{2^m} \otimes \mathbb{I}) c-\UU^{2^m} (QFT_{2^m} \otimes \mathbb{I}) \ket{0} \ket{\psi_k}  .
 \end{eqnarray}

 Note that since the $c-\UU^{2^m}$ no longer depends on the value of the first register, we can commute it through the QFT's to get
  \begin{eqnarray}
 c-\UU^{2^m}  (QFT^{-1}_{2^m} \otimes \mathbb{I})(QFT_{2^m} \otimes \mathbb{I}) \ket{0} \ket{\psi_k}   \nonumber   \\
 =  c-\UU^{2^m} \ket{0} \ket{\psi_k} = e^{2 \pi i 2^m 
\lambda_k} \ket{0} \ket{\psi_k}.
 \end{eqnarray}

This phase factor of  $e^{2 \pi i 2^m \lambda_k}$ is in general a problem when the target register is in a superposition of eigenstates, since different 
eigenstates will pick up a different phase factor.

In some cases, this is not a problem.
One observation, is that such a transformation is equivalent to applying $\UU^{2^m}$. Thus, if our goal is to apply $\UU^{t}$ for $t = \lfloor t \rfloor + 
\delta$, $0 \leq \delta < 1$, where $t \geq 2^m$, we can apply our algorithm for implementing $\UU^{\delta}$, but instead with the above modification. 
This yields an approximation to $\UU^{2^m + \delta}$ with error in $O(\frac{1}{2^m})$. We can then apply $\UU^{t - 2^m}$ to complete the approximation of 
$\UU^t$ with error in $O(\frac{1}{t}) \subseteq O(\frac{1}{2^m})$.

Also note that, if $\lambda_k = \frac{\ell_k}{2^m}$ for an integer $\ell_k$, then the phase factors $e^{2 \pi i 2^m \lambda_k}=1$ all equal $1$, and thus pose 
no problem.

This problem can also be remedied in any other cases where $\lambda_k$ can be determined exactly (or more significantly precisely than error $\frac{1}{2^m}$), 
since before uncomputing $\lambda_k$, one can add an additional phase shift of $e^{-2 \pi i 2^m \lambda_k}$ conditional on the value of $\lambda_k$. This 
will eliminate the final unwanted phase factor of
$e^{2 \pi i 2^m \lambda_k}$ associated with the eigenvector $\ket{\psi_k}$.

\section{Coping with no gap assumption} \label{app:disc}

If one is only interested in the average-case performance ({\it e.g.}, averaging over all possible input states to $\UU^{\frac{1}{2}}$ according to the Haar 
measure), one is still faced with the potential problem that most or all of the spectrum of $\UU$ is in the difficult region (see section \ref{optimality}) of $e^{i \phi} \approx 1$, $\pi < \phi \leq 2 \pi$.

If one is content with obtaining a square root of $e^{i \theta} \UU$ for a random $\theta \in [0, 2 \pi)$, then one can show that the average-case error 
(for any input distribution) is still in $O(\frac{1}{2^m})$ for an algorithm that only uses the square root of $e^{i \theta} \UU$ once.

For an algorithm that uses $(e^{i \theta} \UU)^{\frac{1}{2}}$ a total of $k$ times, the error could get magnified to $O(\frac{k^2}{2^m})$.  Thus, one 
could select $m^{\prime} = m + \lceil 2 \log k \rceil$ for each simulation of $(e^{i \theta} \UU)^{\frac{1}{2}}$.

In order to see why we have a $k^2$ factor instead of $k$, we present an algorithm where substituting an exact square root with our approximate square root yields an error in $\Omega(\frac{k^2}{2^m})$.
The value $\theta$ is of the form $2 \pi \ell/2^m-\epsilon$, for some integer $\ell$ and non-negativeÊvalue $\epsilon \leq 1/2^{m}$. ÊConsider an operator $\UU$ that has everyÊ$2^m$th root of unity $e^{2 \pi i \ell/2^m}$ as an eigenvalue with equal multiplicities.ÊConsider anÊalgorithm that does eigenvalue estimation on the identity operator withÊprecision in $\Theta(1/2^m)$, and uses quantum searching to find anÊeigenvector with eigenvalue $-1$.  Clearly, this algorithm should not find such an eigenvalue since all the eigenvalues are $+1$. Now let $U_1 = (e^{i \theta} \UU)^{-\frac{1}{2}}$ and let $U_2$ (for simplicity, let us assume it is unitary, even though in practice it will be a map that is almost unitary) be our approximation to
$(e^{i \theta} \UU)^{\frac{1}{2}}$ for the same $\theta$.
 Note that $U_1 (e^{i \theta} \UU)^{\frac{1}{2}} = I$, but $U_1 U_2$ will have most eigenvalues near $+1$, and proportion $1/2^m$ of the eigenvalues near $-1$.  Thus, a generalized version of quantum searching (see~\ref{app:genqsearch})ÊwillÊgenerate these (close to) $-1$ eigenvectors with an amplitude inÊ$\Theta(\frac{k}{\sqrt{2^m}})$ using only $k$ evaluations of $U_1 U_2$. ÊRecall that with the ideal $(e^{i \theta} \UU)^{\frac{1}{2}}$ (insteadÊof $U_2$), there would be no such eigenvectors for the search algorithm toÊfind, thus the part of the wave function that found these $-1$ eigenvaluesÊis an error. ÊThis implies an error with probability inÊ$\Theta(\frac{k^2}{2^m})$.

\section{A generalization of quantum search}
\label{app:genqsearch}

Suppose we are given a black box $O_{\phi}$ that flags an unknown quantum state $\ket{\phi} \in H_N$. In other words, $O_{\phi} \ket{\phi} \ket{b} = 
\ket{\phi} \ket{b \oplus 1}$ and $O_{\phi} \ket{\phi^{\prime}} \ket{b} = \ket{\phi^{\prime}} \ket{b}$ for any $\ket{\phi^{\prime}}$ orthogonal to 
$\ket{\phi}$.
For brevity, let us directly use the phase flip version of this black box $O_{\phi} \ket{\phi}  = - \ket{\phi}$ and $O_{\phi} \ket{\phi^{\prime}} =
\ket{\phi^{\prime}}$ for $\bra{\phi} \phi^{\prime} \rangle = 0$ (see Section 8.1 of~\cite{KLM} for a more detailed discussion of the relationship between these black boxes).

Note that if we have a unitary operation $A$ that maps $\ket{00 \ldots 0} \mapsto \sin(\theta) \ket{\phi} + \cos(\theta) \ket{\phi^{\prime}}$ where 
$\bra{\phi} \phi^{\prime} \rangle = 0$, then we can define the slight generalization of a quantum search iterate, $Q = - A U_{00 \ldots 0} A^{-1} 
O_{\phi}$, where $U_{00 \ldots 0} = I - 2 \ket{00 \ldots 0}\bra{00 \ldots 0}$ and note that $Q^k \ket{00 \ldots 0} =  \sin((2k+1)\theta) \ket{\phi} + 
\cos((2k+1)\theta) \ket{\phi^{\prime}}$.  Thus, if we have an algorithm $A$ that ``guesses'' $\ket{\phi}$ with probability $p = \sin^2(\theta)$, then after 
$O(\frac{1}{\sqrt{p}})$ calls to $Q$, we obtain $\ket{\phi}$ with probability very close to $1$.

In traditional quantum searching, the state $\ket{\phi}$ is known to be a computational basis state, and thus it is very easy to produce a unitary 
$A$ such that $|\bra{\phi} A \ket{00 \ldots 0}| = \frac{1}{\sqrt{N}}$, in particular, any operator that maps $\ket{00 \ldots 0} \mapsto \sum_x 
\frac{1}{\sqrt{N}} \ket{x}$ will do.

However, for an arbitrary quantum state $\ket{\phi}$, it is non-trivial to produce such an initial state. One cannot simply use a maximally mixed state, 
since amplitude amplification requires the inverse of the operation $A$ that generates the ``guess''. We could resort to techniques for generating pure 
states that for practical purposes are ``random'' ({\it e.g.}, \cite{AE07}),
however, for this particular application, it is sufficient to use the following idea.

Let $A$ act on an $N^2$ dimensional Hilbert space and maps $\ket{00 \ldots 0} \ket{00 \ldots 0} \mapsto \sum_x \frac{1}{\sqrt{N}} \ket{x}\ket{x}$. Note 
that for any basis $\{ \ket{\phi_j} \}$, this maximally mixed state can be rewritten as $ \sum_j \frac{1}{\sqrt{N}} \ket{\phi_j}\ket{\phi_j^*}$ (where the 
coefficients of $\ket{\phi_j^*}$ are the complex conjugates of those of $\ket{\phi}$).  Notice that the maximally entangled state has equal support for every eigenvector in the basis of eigenvalues of the black box unitary, which we require for
the procedure outlined in this section to succeed.  Let us assume, without loss of generality, that $\ket{\phi_0} = \ket{\phi}$.
If we apply amplitude amplification, using $Q = - A U_{00 \ldots 0, 0 \ldots 0} A^{-1} (O_{\phi} \otimes I)$, then it is easy to verify that
\begin{eqnarray} 
Q^k \ket{00 \ldots 0} \ket{00 \ldots 0} &=& \sin((2k+1)\theta) \ket{\phi_0} \ket{\phi^*_0}     \nonumber  \\
& & + \cos((2k+1)\theta) \left( \sum_{j=1}^{N-1} \frac{1}{\sqrt{N-1}} \ket{\phi_j} \ket{\phi_j^*} \right),  
\end{eqnarray}
where $\sin(\theta) = \frac{1}{\sqrt{N}}$.  Thus, by choosing $k \approx \frac{\pi}{4 \theta} \in O(\frac{1}{\sqrt{p}})$, we obtain $\ket{\phi_0} 
\ket{\phi_0^{*}}$ with high fidelity.  For applications where the extra $\ket{\phi^{*}}$ system will pose a problem, this approach will not work, and 
techniques for generating pure states that behave like random quantum states could be used instead.

This technique also works if the black box flags a subspace of dimension $d$.
Again, assume without loss of generality that this subspace is spanned by $\ket{\phi_0}, \ket{\phi_1}, \ldots, \ket{\phi_{d-1}}$.
Then the maximally entangled state can be rewritten as
\begin{equation}
\sin(\theta) \left( \sum_{j=0}^{d-1} \frac{1}{\sqrt{d}} \ket{\phi_j} \ket{\phi_j^*} \right) + \cos(\theta) \left( \sum_{j=d}^{N-1} \frac{1}{\sqrt{N-d}} 
\ket{\phi_j} \ket{\phi_j^*} \right)
\end{equation}
where $\sin^2(\theta) = \frac{d}{N}$.

Thus $Q^k \ket{00 \ldots 0}\ket{00 \ldots 0} \approx \sum_{j=0}^{d-1} \frac{1}{\sqrt{d}} \ket{\phi_j} \ket{\phi_j^*}$ for $k \approx \frac{\pi}{4} 
\sqrt{\frac{N}{d}}$, assuming $\sqrt{\frac{d}{N}}$ is much less than one.

Furthermore, if we applied amplitude estimation~\cite{BHMT} in this case, we would be approximating the value of $\theta = \arcsin(\sqrt{\frac{d}{N}})$, which would allow us to approximate the dimension of the subspace, $d$.

\end{document}